\documentclass[11pt,a4paper]{article}
\usepackage{graphics}

\begin{document}
\baselineskip=14pt

\begin{center}

{\Large\bf
Influence of lasers propagation delay on the sensitivity of atom
interferometers }

\bigskip{
J. Le Gou\"{e}t, P. Cheinet, J. Kim, D. Holleville,  A. Clairon,
A. Landragin, F. Pereira Dos Santos }

\smallskip{\small{\it
LNE-SYRTE, CNRS UMR 8630 \\ Observatoire de Paris, 61 av. de
l'Observatoire, 75014 Paris, France }

E-mail : franck.pereira@obspm.fr

}\end{center}

\medskip

\abstract{In atom interferometers based on two photon transitions,
the delay induced by the difference of the laser beams paths makes
the interferometer sensitive to the fluctuations of the frequency
of the lasers. We first study, in the general case, how the laser
frequency noise affects the performance of the interferometer
measurement. Our calculations are compared with the measurements
performed on our cold atom gravimeter based on stimulated Raman
transitions. We finally extend this study to the case of cold atom
gradiometers.}

%
\medbreak
\section{Introduction}
\label{intro} Atom interferometry allows to realize measurements
in the fields of frequency metrology \cite{Santarelli}, inertial
sensors \cite{Kasevich,Riehle}, tests of fundamental physics
\cite{Weiss,Marion,Fred}. This technique is based on the splitting
of an atomic wave function into separated wave packets. The
difference in the quantum phases accumulated by the wave packets
can be extracted from the interference pattern obtained when
recombining them. Among the various types of coherent beam
splitters developed for matter wave manipulation
\cite{Keith,Rasel,Miniatura,Schumm,Berman}, two photon transitions
have proven to be powerful tools for precise measurements. For
instance, atom interferometers based on Bragg transitions
\cite{Rasel} can be used for polarisability \cite{Miffre} and
fundamental measurements \cite{Muller}. Stimulated Raman
transitions \cite{Moler92} allowed the development of high
precision inertial sensors \cite{Chu,Gustavson,Snadden,Canuel},
whose performances compete with state of the art instruments
\cite{Faller,Stedman}.

In the case of interferometers based on two photon transitions,
atomic wave packets are split and recombined with light pulses of
a pair of counter-propagating laser beams, which couple long lived
atomic states. The sensitivity of such interferometers arises from
the large momentum transfer of counter-propagating photons. A
propagation delay is unavoidable between the two
counter-propagating beams at the position of the atoms, and we
show here that this delay makes the interferometer measurement
sensitive to the lasers frequency noise. Without losing
generality, we detail this effect in the case of our gravimeter,
based on stimulated Raman transitions. However, the formalism
presented here can be applied to any type of interferometer where
two photon transitions are used as beam splitters.

The sensitivity to inertial forces of such an interferometer
arises from the imprinting of the phase difference between the
lasers onto the atomic wave function \cite{borde89}. As temporal
fluctuations in this laser phase difference affect the measurement
of the atomic phase, a high degree of phase coherence is required.
This coherence can be obtained either by using two sidebands of a
single phase modulated laser \cite{Kasevich}, or by locking the
phase difference between two independent lasers
\cite{Santarelli2,Bouyer}. In both cases, the phase relation is
well determined only at a specific position, where the laser is
modulated or where the frequency difference is measured. Between
this very position and the atoms, this phase difference will be
affected by fluctuations of the respective paths of the two beams
over the propagation distance. In most of the high sensitivity
atom interferometers, the influence of path length variations is
minimized by overlapping the two beams, and making them propagate
as long as possible over the same path. The vibrations of any
optical element shift the phase of each laser, but do not strongly
disturb their phase difference as long as the lasers co-propagate,
because their optical frequencies are very close. However, for the
interferometer to be sensitive to inertial forces, the two beams
(with wave vectors $\vec{k}_1$ and $\vec{k}_2$) have to be
counter-propagating. The two overlapped beams are thus directed to
the atoms and retro-reflected. Among the four beams actually sent
onto the atoms, two will realize the interferometer pulses. As a
consequence, the reflected beam is delayed with respect to the
other one. The phase difference at the atoms position is then
affected by the phase noise of the lasers, accumulated during this
reflection delay.

In this article, we investigate both theoretically and
experimentally the influence of the delay on the sensitivity of an
atom interferometer. In the following section, we briefly describe
our experimental setup. The transfer function of the
interferometer phase noise with respect to the Raman laser
frequency noise is derived in section 3, and compared with
experimental measurements. In section 4, we demonstrate the
sensitivity limitations induced by the retro-reflection delay of
the lasers in the case of our atomic gravimeter. We then discuss
how such limitations could be overcome. The discussion is finally
extended to the case of high precision gradiometers, whose
performances might be limited by their intrinsic propagation
delays.

\section{Experimental setup}
\label{Exp setup}

Our interferometer is a cold atom gravimeter based on stimulated
Raman transitions, which address the two hyperfine sublevels $F=1$
and $F=2$ of the $^5S_{1/2}$ ground state of the $^{87}$Rb atom.
We use successively a 2D-MOT, a 3D-MOT and an optical molasses to
prepare about $10^7$ atoms at a temperature of $2.5~\mu
\textrm{K}$, within a loading time of 50~ms. The intensity of the
lasers is then adiabatically decreased to drop the atoms, and we
detune both the repumper and cooling lasers from the atomic
transitions by about 1~GHz to obtain the two off-resonant Raman
lasers. A description of the compact and agile laser system that
we developed can be found in \cite{PatocheAPB}. The preparation
sequence ends with the selection of a narrow velocity distribution
($\sigma_v \leq v_r = 5.9 \textrm{mm/s}$) in the $\left|F=1,
m_F=0\right\rangle$ state, using a combination of microwave and
optical pulses.

A sequence of three pulses ($\pi/2-\pi-\pi/2$) then splits,
redirects and recombines the atomic wave packets. At the output of
the interferometer, the transition probability from an hyperfine
state to the other is given by the usual formula of two waves
interferometers : $P=\frac{1}{2}\left(1 + C\cos\Delta\Phi\right)$,
where $C$ is the contrast of the fringes, and $\Delta\Phi$ the
difference of the atomic phases accumulated along the two paths.
We measure by fluorescence the populations of each of the two
states and deduce the transition probability. The difference in
the phases accumulated along the two paths depends on the
acceleration $\vec{a}$ experienced by the atoms. It can be written
as
$\Delta\Phi=\phi(0)-2\phi(T)+\phi(2T)=\vec{k}_{eff}.\vec{a}T^{2}$,
where $\phi(0,T,2T)$ is the phase difference of the lasers at the
location of the center of the atomic wavepackets for each of the
three pulses \cite{borde02},
$\vec{k}_{eff}=\vec{k}_{1}-\vec{k}_{2}$ is the effective wave
vector (with $|\vec{k}_{eff}|=k_1+k_2$), and $T$ is the time
interval between two consecutive pulses \cite{Kasevich}.

\begin{figure}
    \resizebox{0.75\columnwidth}{!}{
        \includegraphics{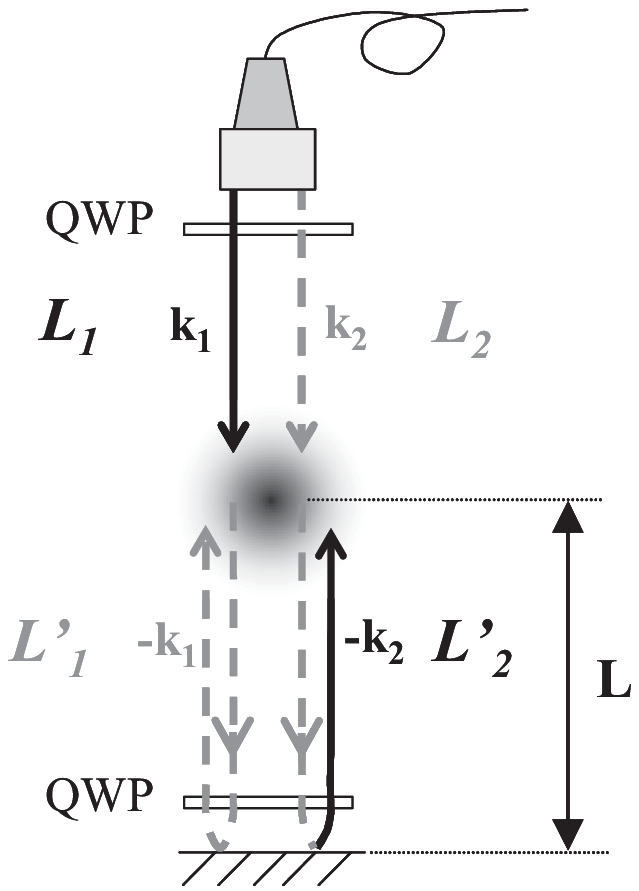}}
        \caption{Experimental scheme of the cold atom gravimeter. The two Raman lasers $L_1$ and $L_2$ are guided from the optical bench
        to the atoms by the same optical fiber, and the resonant counter-propagating beams are obtained by retro-reflecting the lasers with
         the mirror at the bottom of the vacuum chamber. Due to the Doppler shift of the falling atoms, only $L_1$ and $L'_2$ can drive the
         Raman transitions. QWP: quarter wave plate.}
        \label{Image1}
\end{figure}

The Raman light sources are two extended cavity diode lasers,
amplified by two independent tapered amplifiers. Their frequency
difference is phase locked onto a microwave reference source
generated by multiplications of highly stable quartz oscillators.
The two Raman laser beams are overlapped with a polarization beam
splitter cube, resulting in two orthogonally polarized beams.
First, a small part of the overlapped beams is sent onto a fast
photodetector to measure an optical beat. This beat-note is first
mixed down with a reference microwave oscillator, and finally
compared to a stable reference RF frequency in a Digital Phase
Frequency Detector. The phase error signal is then used to lock
the laser phase difference at the very position where the beat is
recorded. The phase locked loop reacts onto the supply current of
one of the two lasers (the "slave" laser), as well as on the
piezo-electric transducer that controls the length of its extended
cavity. The impact of the phase noise of the reference microwave
oscillator on the interferometer sensitivity, as well as the
performances of the PLL has already been studied in
\cite{PatrickIEEE}. Finally, the two overlapped beams are injected
in a polarization maintaining fiber, and guided towards the vacuum
chamber. We obtain the counter-propagating beams by laying a
mirror and a quarterwave plate at the bottom of the experiment. As
displayed in figure \ref{Image1}, four beams ($L_1$, $L_2$,
$L'_1$, $L'_2$) are actually sent onto the atoms. Because of the
selection rules and the Doppler shift induced by the free fall of
the atoms, only the counter-propagating pair $L_1$/$L'_2$ drives
the Raman transitions. In the following, we define $L_1$ as the
"master" laser, and $L_2$ as the "slave" one.

\section{Influence of the propagation delay on the interferometer phase noise}
\label{phase_diff}

\subsection{Theoretical expression of the transfer function}
\label{par_3_1}

The phase difference $\varphi$ imprinted onto the atoms by the
counter-propagating beams is given by
$\varphi(t)=\varphi_{1}(t)-\varphi_{2'}(t)$, where $\varphi_1$ and
$\varphi_{2'}$ are respectively the phases of the
downward-propagating master laser and of the retro-reflected slave
laser. Because of the retro-reflection, the  phase of $L'_2$
writes as $\varphi_{2'}(t)=\varphi_{2}(t-t_d)$. The
retro-reflection delay $t_d$ is given by $t_{d}=2L/c$, where $L$
is the distance between the atoms and the bottom mirror. We
consider here a perfect phase locked loop, which guaranties the
stability of the phase difference for copropagating lasers. Then
$\varphi_{2}(t-t_d)=\varphi_{1}(t-t_d)+\omega_0 \times (t-t_d)$,
where $\omega_0$ is the frequency difference between the two
lasers. Since we assume $\omega_0$ is perfectly stable, its
contribution will vanish in the interferometer phase $\Delta\Phi$.
Thus, we do not take it into account when writing the laser phase
difference and finally obtain
$\varphi(t)=\varphi_{1}(t)-\varphi_{1}(t-t_d)$.

As shown in \cite{PatrickIEEE}, the interferometer phase shift
$\Phi$ induced by fluctuations of $\varphi$ can be written as:
\begin{equation}
\label{sphi}
\Phi=\int^{+\infty}_{-\infty}{g(t)\frac{d\varphi(t)}{dt}dt}
\end{equation}
where $g(t)$ is the sensitivity function of the interferometer.
This function quantifies the influence of a relative laser phase
shift $\delta\phi$ occurring at time t onto the transition
probability $\delta P(\delta\phi,t)$. It is defined in \cite{Dick}
as:
\begin{equation} \label{g(t)}
g(t)=2\lim_{\delta\phi\rightarrow 0}\frac{\delta
P(\delta\phi,t)}{\delta\phi}
\end{equation}
We consider an interferometer with three pulses $\pi/2-\pi-\pi/2$
of durations respectively $\tau_R - 2\tau_R - \tau_R$. If the time
origin is chosen at the center of the $\pi$ pulse, $t \mapsto
g(t)$ is an odd function. Its following expression for positive
time is derived in \cite{PatrickIEEE}:
\begin{equation} \label{g_entier}
g(t) = \left\{\begin{array}{ll}
            \sin \Omega_R t & \textrm{for $0<t<\tau_R$}\\
            1 & \textrm{for $\tau_R<t<T+\tau_R$}\\
            -\sin \Omega_R (T - t) & \textrm{for $T+ \tau_R<t<T+2\tau_R$}\\
            \end{array} \right.
\end{equation}
where $\Omega_R$ is the Rabi frequency.

In the presence of fluctuations of the master Raman laser
frequency, the interferometer phase shift becomes:
\begin{eqnarray}
    \Phi &=& \int^{+\infty}_{-\infty}{dt\:g(t)\frac{d\varphi(t)}{dt}} \nonumber \\
         &=& \int^{+\infty}_{-\infty}{dt\:g(t)\left[\frac{d\varphi_1(t)}{dt}-\frac{d\varphi_1(t-t_d)}{dt} \right]}
\end{eqnarray}
If no assumption is made on the distance $L$ between the mirror
and the atoms, the retro-reflection delay $t_d$ is not the same
for the three pulses. However, in our experiment, the maximum
duration of an interferometer is 100~ms, which corresponds to a
5~cm atomic path, much smaller than the distance $L \approx
50$~cm. We can thus consider $t_d$ constant during the
measurement, and write the interferometer phase shift as:
\begin{eqnarray}\label{phi2}
    \Phi &=& \int^{+\infty}_{-\infty}{dt\:\left[g(t)-g(t+t_d)\right]\frac{d\varphi_1(t)}{dt}} \nonumber \\
         &=& \int^{+\infty}_{-\infty}{dt\:\left[g(t)-g(t+t_d)\right]\nu_1(t){dt}}
\end{eqnarray}
We deduce from (\ref{phi2}) that the transfer function $Z$, which
converts Raman laser frequency noise into interferometer phase
noise, is given by the Fourier transform of the difference
$g(t)-g(t+t_d)$. After some algebra, we find:
\begin{equation} \label{dphi_modul}
Z(f,t_d) = -i e^{-i\omega t_{d}/2} \times t_d \times H(2\pi f)
\times \frac{\sin\left(\pi ft_d\right)}{\pi f t_d}
\end{equation}
where $\displaystyle H(\omega)=\omega\int{g(t)e^{i\omega t}dt}$ is
the weighting function describing the response of the
interferometer phase to the fluctuations of the laser phase
difference, as already described in \cite{PatrickIEEE}. A
remarkable feature of the function $H(\omega)$ is a low pass first
order filtering, arising from the fact that the response time of
the atoms to a perturbation is necessarily limited by the Rabi
frequency. The cutoff frequency is given by
$f_c=\sqrt{3}\Omega_R/6\pi=\sqrt{3}/12\tau_R$.

In our experimental setup, the delay time is about $t_d=3$~ns. Since the cut-off frequency $f_c$ is roughly 20~kHz, we can assume that $f_c t_d \ll 1$. 
The amplitude of the transfer function is finally:
\begin{equation} \label{fct_transfert}
\left|Z(f,t_d)\right| \approx t_d \: \left|H(2\pi f)\right|.
\end{equation}

\subsection{Measurement of the transfer function}
\label{par_3_2} In order to measure the amplitude of $Z(f)$, we
modulate the master laser frequency at a frequency $f$. The
applied frequency modulation is detected in the beat-note between
the master laser and a 'reference' laser, locked on a atomic line
of the $^{87}$Rb by a saturated spectroscopy setup. The frequency
of the beat-note is converted into a voltage modulation by a
frequency to voltage converter (FVC). When the modulation is not
synchronous with the cycle rate, the response of the
interferometer appears as a periodic modulation of its phase. Its
amplitude is the modulus of the transfer function, and the
apparent period of the response depends on the ratio $f/f_s$,
where $f_s$ is the sampling rate of the experiment. For these
measurements, the cycle rate was $f_s=4$~Hz.

We choose the modulation frequency as $f=(n+1/10)f_s$ and record
the transition probability from which we extract the transfer
function amplitude $\left|Z(f,t_d)\right|$. We run the experiment
with a modest interrogation time of $2T=2$~ms, which allows us to
reach a good signal to noise ratio (SNR) of 250 per shot for the
detection of the perturbation. As the interferometer phase shift
scales as the square of $T$, best sensitivities to inertial forces
are usually obtained for large values of $T$. However, in that
case, the interferometer also becomes more sensitive to
vibrations, which limit the SNR to about 50 in our experiment when
$2T = 100$~ms.

\begin{figure}
    \resizebox{1\columnwidth}{!}{
        \includegraphics{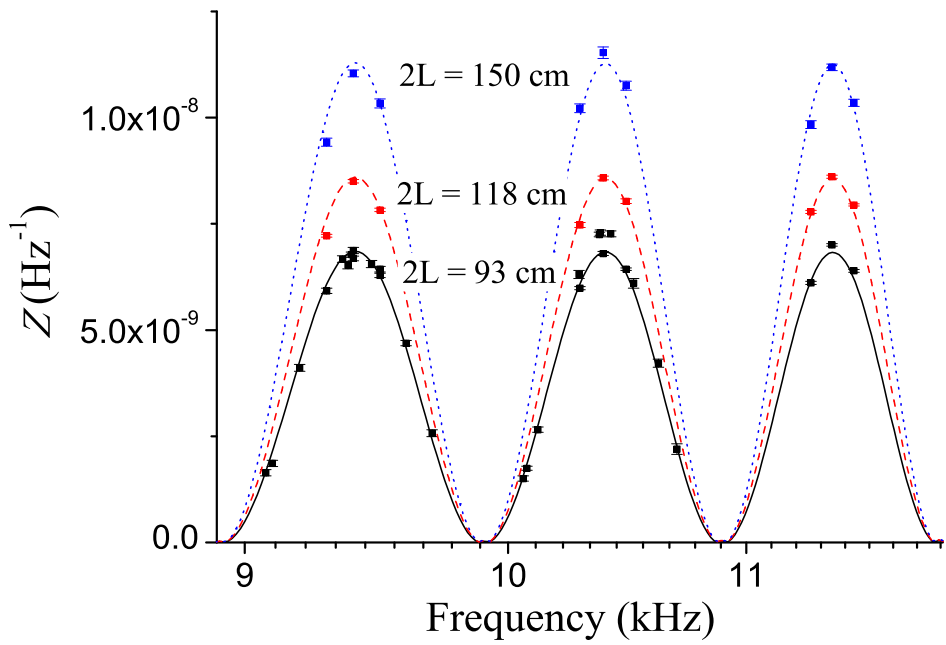}}
    \caption{Transfer function $Z$ of the frequency noise of the laser for three optical lengths. The experimental points and the theoretical
    curves (see equation (\ref{fct_transfert})) are in good agreement.}
    \label{graph2}
\end{figure}

Figure \ref{graph2} displays the measured and calculated transfer
function $Z$ as a function of the modulation frequency $f$, for
three values of the retro-reflection length: $2L$~=~93, 118 and
150~cm. The weighting function zeros occur when the period of the
perturbation is a multiple of $T+2\tau_R$. In that case, the phase
of the perturbation is the same for each of the three pulses, and
the corresponding interferometer phase shift
$\Delta\Phi=\varphi_1-2\varphi_2+\varphi_3$ vanishes. One can see
on figure \ref{graph2} that the experimental points agree with the
calculation (eq. \ref{fct_transfert}), demonstrating that the
amplitude of $Z$ increases linearly with the time delay $t_d$.

\begin{figure}
    \resizebox{1\columnwidth}{!}{
        \includegraphics{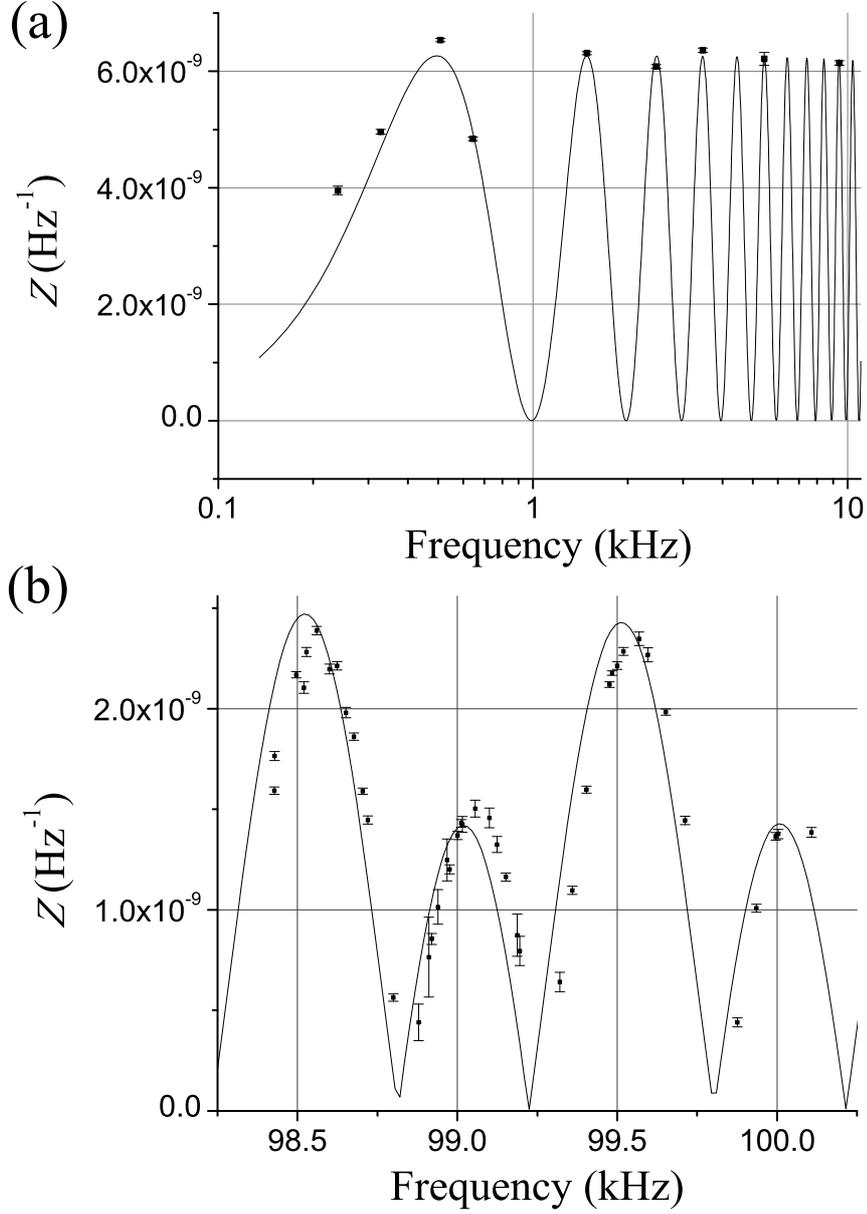}}
    \caption{Calculation and measurement of the transfer function for low (a) and high (b) frequencies (with respect to $f_c \approx 24 kHz$) of
    master frequency modulation. For these measurements, the back and forth distance between the atoms and the mirror is $2L=93$~cm.}
    \label{graph3}
\end{figure}

We also test further the relation between our measurement of the
transfer function and the weighting function $H(\omega)$
\cite{PatrickIEEE}. We measure the transfer function for a fixed
value of $t_d$, for frequencies respectively lower and higher than
the low pass cut-off frequency $f_c$. In our case, a $\pi/2$ pulse
is $6\:\mu s$ long, so $f_c$ is about 24~kHz. The measurements are
presented in figure \ref{graph3}. For $f \gg f_c$, there is a
slight shift between the measurement and the theoretical
expression of $Z$. We tested out various possible origins like the
duration and timings of the pulses, the synchronization of the
frequency synthesizer we used to modulate the laser frequency and
the clock frequency of the experiment, but this shift is still not
understood. However it does not affect the value of the variance
integrated over the whole spectrum (see eq. \ref{var_approx}).

\section{Limits on the interferometer sensitivity}
\label{par_4}

\subsection{Theoretical analysis}
\label{par_4_1}

We finally quantify the degradation of the interferometer
sensitivity as a function of the laser frequency noise level and
of the optical delay. Using the equation (\ref{phi2}), the
variance of the phase fluctuation is given by:

\begin{equation} \label{var1}
\sigma_{\Phi}^{2} = \int^{+\infty}_{0} {\left|Z(\omega)\right|^{2}
S_{\nu_{1}}(\omega) \frac{d\omega}{2\pi}}
\end{equation}
where $S_{\nu_{1}}$ is the power spectral density (PSD) of the
master laser frequency noise. Then, using equation
(\ref{dphi_modul}), one writes the variance as:

\begin{equation} \label{var3}
\sigma^{2}_{\Phi}=4 \int^{+\infty}_{0} {\sin^2\left(\frac{\omega
t_{d}}{2}\right) \frac{\left|H(\omega)\right|^{2}}{\omega^2}
S_{\nu_{1}}(\omega) \frac{d\omega}{2\pi}}
\end{equation}
The same approximation than before ($\pi ft_{d} \ll 1$) leads to
the final expression:

\begin{equation} \label{var_approx}
\sigma^{2}_{\Phi} \approx t_{d}^{2} \int^{+\infty}_{0}
{\left|H(\omega)\right|^{2} S_{\nu_{1}}(\omega)
\frac{d\omega}{2\pi}}
\end{equation}

According to this formula, the interferometer sensitivity
$\sigma_{\Phi}$ increases linearly with the retro-reflection
length. In the case of a white frequency noise
($S_{\nu_1}(\omega)=S_{\nu_1}^0$), the variance is:
\begin{equation} \label{var_white}
\sigma^{2}_{\Phi} \approx \frac{\pi^2}{4\tau_R} \: t_{d}^{2} \:
S_{\nu_1}^0
\end{equation}
This last result gives a simple evaluation of the level of white
frequency noise required to reach a given sensitivity, for given
retro-reflection delay and Raman pulse duration.

\subsection{Example of the laser frequency noise influence}
\label{par_4_2}

\begin{figure}
    \resizebox{1\columnwidth}{!}{
        \includegraphics{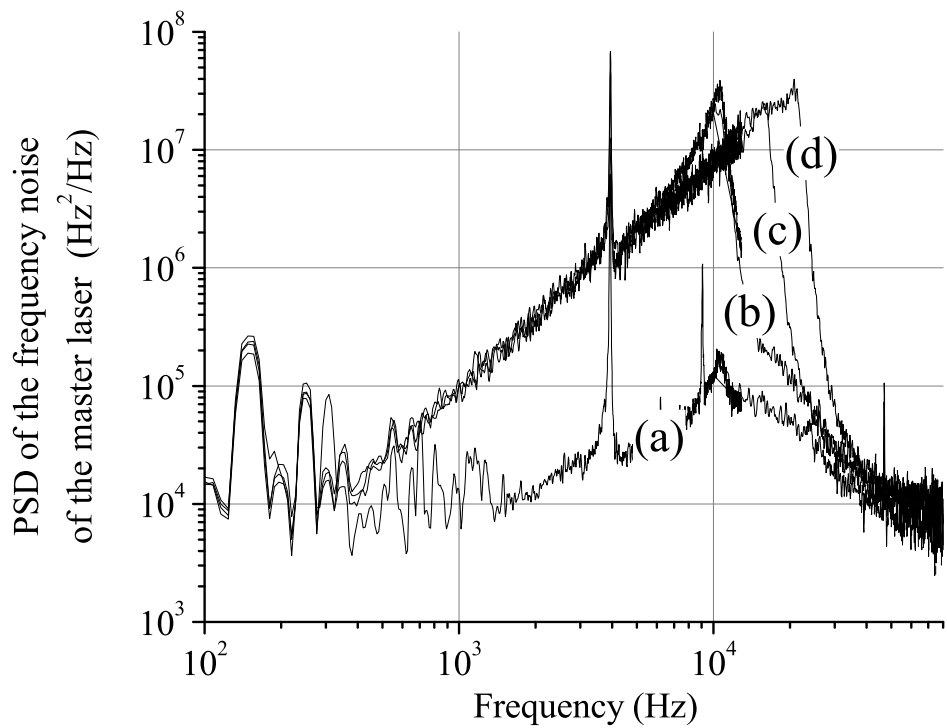}}
    \caption{PSD of the frequency noise of the master laser. The curve (a) shows a typical unperturbed power spectrum of the
laser. The other curves correspond to the PSD with added noise on
the laser current, for different cut-off frequencies of the low
pass filter  : (b) 10 kHz, (c) 15 kHz, (d) 20 kHz.}
    \label{graph4}
\end{figure}

\begin{figure}
    \resizebox{1\columnwidth}{!}{
        \includegraphics{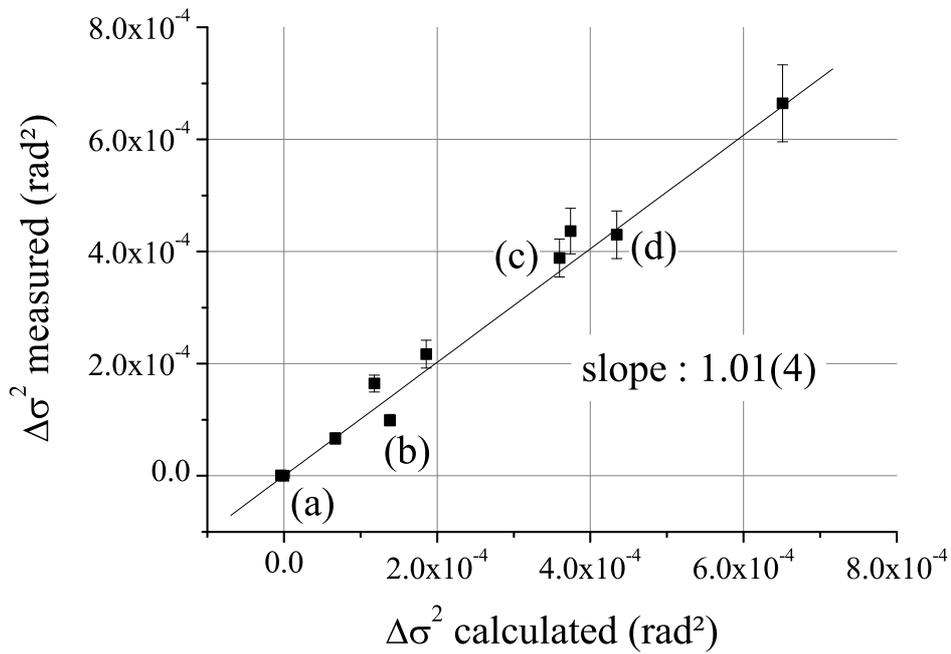}}
    \caption{Comparison between calculated and measured degradations of the phase sensitivity, for different added noise. The point (a),
    where $\Delta \sigma^{2}=0$, corresponds to the case where no frequency noise is added. The points (b), (c) and (d) correspond
    to the power spectra displayed in figure \ref{graph4}.}
    \label{graph5}
\end{figure}

In a second experiment, the frequency noise is deliberately
degraded by adding noise on the master laser current. We use a
high gain amplifier with an incorporated tunable low pass filter
(Stanford Research System SR650) as the noise source, with its
input connected to ground. We basically control the amount of RMS
frequency noise by changing the cut-off frequency of the filter
(see fig. \ref{graph4}). The PSD of the master laser frequency
noise is measured by analyzing the FVC output with a FFT analyzer
(we made sure it is well above the PSD of the reference laser to
which the master laser is compared). We also measure the power
spectrum of the laser without additional noise, and we calculate
the two corresponding variances, with or without added noise,
using equation (\ref{var_approx}). The difference between the two
variances gives the expected variance degradation
$\Delta\sigma^2_{\Phi}$ of the interferometer phase noise. We
compare this calculation with the experimental value of
$\Delta\sigma^2_{\Phi}$ obtained by measuring the difference
between the variances of the interferometer phase with and without
added noise. The experiment was performed for $2L= 93$ cm, and the
figure \ref{graph5} shows the comparison between the calculated
and the measured values of the variance degradation. The
experimental values agree very well with the result of the
calculation.

From the nominal frequency noise spectrum (curve \textit{(a)} on
figure \ref{graph4}), we estimate that the retro-reflection
induces a laser frequency noise contribution of 2.4~mrad/shot to
the total interferometer noise.

\section{Discussion}
\label{par_5}

\subsection{Sensitivity limitation of the gravimeter measurement}
\label{par_5_1}

This contribution of the frequency noise does not depend on the
duration $2T$ of our interferometer. Indeed, as discussed before,
the retro-reflection delay $t_d$ can be considered as constant
even for the longest interferometer we can perform. Moreover,
dominant contributions to the variance arise from the high
frequency part of the laser frequency noise spectrum, for which
the fast oscillations of the transfer function average to the same
value, regardless to $2T$.

The calculated laser frequency noise contribution induced by the
retro-reflection is of the same order of magnitude than the other
sources of phase noise also due to the lasers. Indeed, the PLL
noise contributes for 2.1 mrad/shot \cite{PatrickIEEE}, the
various frequency references for 1.6 mrad/shot \cite{PatocheAPB},
and the propagation in the optical fiber for 1.0 mrad/shot. All
these noise sources are independent, so the frequency noise of the
Raman lasers represents a total contribution of
$\sigma_{\Phi}=3.7$ mrad/shot to the interferometer phase
sensitivity.

The phase sensitivity of 2.4 mrad/shot limits the sensitivity of
the acceleration measurement up to $\sigma_g=3\times 10^{-9}
\textrm{g}/\sqrt{\textrm{Hz}}$ with our experimental parameters
($2T=100$ ms, $\tau_R=6$ $\mu s$, $L=93$ cm, and cycle rate 4 Hz).
However, the interferometer sensitivity is presently limited to
$2.10^{-8} \textrm{g}/\sqrt{\textrm{Hz}}$ by the vibration noise.

We want to emphasize here that our ECDL have excellent white
frequency noise floor, which corresponds to a linewidth of only 5
kHz. Excess 1/f noise at low frequency is inherent to the diode
lasers. It could be reduced more efficiently by using other
locking techniques which allow larger bandwidths
\cite{PDH,Dahmani,Crozatier}. Other laser sources based on
frequency doubled fiber lasers, whose frequency noise is extremely
low, could be beneficial \cite{Thompson,ONERA}. On the contrary,
DBR laser diodes, whose linewidth is typically a few MHz, are not
recommended.

According to equation (\ref{var_white}), the sensitivity may be
improved by using longer Raman pulses. On the other hand, when the
duration $\tau_R$ is larger, the velocity selectivity of the
pulses becomes more stringent. Then the contribution of useful
atoms to the signal is smaller, and the detection noise is larger.
Even for the lowest temperatures one can reach with
$\sigma_+-\sigma_-$ cooling, the increase of $\tau_R$ reduces
either the the contrast when no primary velocity selection is
performed, or the number of atoms in the measurement. Ultra-cold
atoms, obtained by evaporative or sideband cooling, would be of
interest \cite{Doyle,Perrin}.

The sensitivity can also be improved by bringing the mirror closer
to the atoms. Presently, our mirror is located at the bottom of
the experiment, out of the magnetic shields. Ultimately the mirror
could be installed inside the vacuum chamber, very close to the
atoms. In this ideal situation, the laser propagation delay cannot
be considered constant for the three pulses anymore. The maximum
delay scales as the trajectory length, which is proportional to
$T^2$. On the other hand, the sensitivity to inertial forces also
scales as $T^2$ when going to large interaction times. Hence, the
sensitivity limit on the inertial measurement induced by the
propagation delay, does not depend on $T$ for ground instruments.
The situation is more favorable for space based instruments
\cite{ONERA} where the distance between the atoms and the
retro-reflection mirror would scale like the separation of the
wavepackets, meaning only like $T$.

\subsection{Influence on gradiometers measurement}
\label{par_5_2}

The formalism developed here could finally be useful to determine
the ultimate performances of cold atom gradiometers. In such
experiments, two atomic clouds are spatially separated and realize
simultaneously gravity measurements \cite{Snadden,Maleki}. Most of
the phase noise contributions are rejected thanks to the
differential measurement, when the clouds experience the same
Raman lasers. However, as the lasers propagation delays are not
the same for the two spaced interferometers, the laser frequency
noise do not cancel. Let us consider the simple case where the
atomic sample $S_2$ is very close to the retro-reflection mirror,
whereas the other $S_1$ is half a meter above. While the phase
noise induced by the laser $L'_2$ propagation is negligible for
$S_2$, for the other sample $S_1$ this phase noise contribution
would reach the 2.4~mrad/shot that we calculated for a single
sample located at $L = 93/2$~cm, with our laser setup. A
remarkable point is that this phase noise contribution scales like
the distance $L=c t_d/2$, just like the sensitivity of the
gradiometer measurement. Hence there would be no advantage in
increasing the separation between the samples, as long as one do
not increase the interaction time $2T$.

\begin{figure}
    \resizebox{.75\columnwidth}{!}{
        \includegraphics{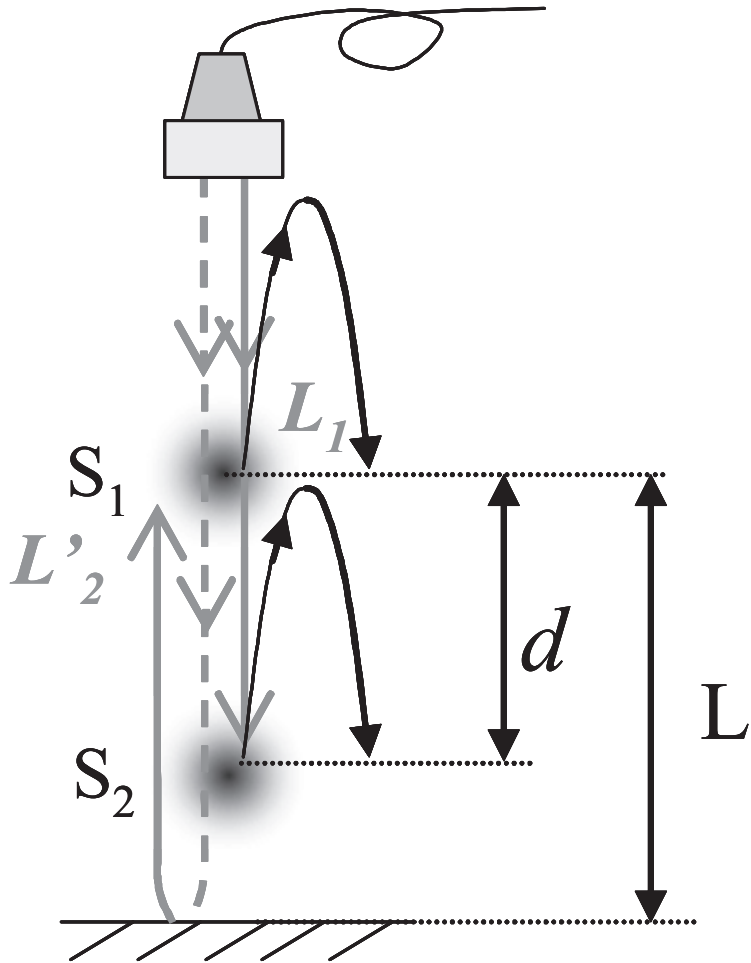}}
    \caption{Possible setup of a cold atom gradiometer, where two samples $S_1$ and $S_2$ are used for two simultaneous interferometers.
    Their separation $d$ keeps constant all along their trajectories, and the phase noise induced by the frequency noise of $L'_2$ during
    the retro-reflection only depends on $d$.}
    \label{Image2}
\end{figure}

In the more common configuration where the samples are given the
same initial velocity, the distance $d$ between them remains
constant during their trajectories. It is then quite
straightforward that the gradiometer phase noise induced by the
lasers propagation delays only depend on the separation $d$. Thus
the sensitivity limit is also given by the equation
\ref{var_approx}, with $t_d = 2d/c$. The variance in the case of a
white frequency noise is then:
\begin{equation} \label{var_gradio}
\sigma^{2}_{\Phi} \approx \frac{\pi^2}{\tau_R} \: \frac{d^2}{c^2}
\: S_{\nu_1}^0
\end{equation}
Using our experimental setup, with the parameters mentioned
before, the best sensitivity would be thus
$60~\textrm{E}/\sqrt{\textrm{Hz}}$ ($1\textrm{E}=10^{-9}~s^{-2})$.
Let us consider now an atomic fountain configuration with a
vertical separation $d = 1$~m of the two samples, and a trajectory
height of 1 meter too (see figure \ref{Image2}). This trajectory
is obtained for an initial velocity of 4~m/s, and the apogee is
reached after a time interval of 450~ms, which defines the
interaction time $T$. A laser linewidth as small as 500~Hz
(corresponding to a white frequency noise of about $S_\nu =
160~\textrm{Hz}^2/\textrm{Hz}$) would allow to obtain a stability
measurement of $1~\textrm{E}/\sqrt{\textrm{Hz}}$ (for a standard
pulse duration $\tau_R = 10~\mu s$).

\section{Conclusion}

We have investigated the influence of the optical propagation
delays on the phase noise of an atom interferometer based on two
photon transitions. The transfer function for the laser frequency
fluctuations has been calculated and measured for various optical
paths with our cold atom gravimeter. Quantitative measurements of
the interferometer sensitivity have also been performed, which
show that the laser frequency noise can limit the sensitivity of
the interferometer. We therefore suggest that a necessary effort
must be placed to reduce the laser frequency noise. Thus for
experiments where vibrations are not the main limitations, for
instance in the case of space applications, integrated DFB or DBR
lasers are not recommended. We apply the present formalism to the
case of atomic gradiometers, where the other sources of
interferometer phase noise are rejected. A model is proposed to
estimate the required frequency laser noise in order to reach a
given sensitivity. This work presents interest for spaceborne
experiments as well, where interaction times can be much longer,
and where the effect of the lasers propagation could constitute a
technical limitation.

The authors would like to thank the Institut Francilien pour la
Recherche sur les Atomes Froids (IFRAF), the Centre National des
Etudes Spatiales (contract no. 02/CNES/0282), the European Union
(FINAQS) for financial support. P.C. and J.L.G. thank DGA for
supporting their works.

%

\newpage

\end{document}